\documentclass[aps,pra,twocolumn,superscriptaddress]{revtex4-2}
\usepackage{amsmath}
\usepackage[colorlinks,linkcolor=blue,anchorcolor=blue,citecolor=blue,]{hyperref}
\usepackage{braket}
\usepackage{color}
\usepackage[capitalise]{cleveref}
\usepackage{threeparttable}
\usepackage{graphicx}

\newcommand{\Apar}{A_{\parallel}}
\newcommand{\Aperp}{A_{\perp}}

\newcommand{\mApar}{A_{\parallel}}
\newcommand{\mAperp}{A_{\perp}}

\begin{document}
\title{Benchmarking of the Fock space coupled cluster method and uncertainty estimation: Magnetic hyperfine interaction in the excited state of BaF}

\author{Malika Denis}
\thanks{These two authors contributed equally}
\affiliation{Van Swinderen Institute for Particle Physics and Gravity, University of Groningen, 9747 AG Groningen, The Netherlands}
\affiliation{Nikhef, National Institute for Subatomic Physics, 1098 XG Amsterdam, The Netherlands}

\author{Pi A. B. Haase}
\thanks{These two authors contributed equally}
\affiliation{Van Swinderen Institute for Particle Physics and Gravity, University of Groningen, 9747 AG Groningen, The Netherlands}
\affiliation{Nikhef, National Institute for Subatomic Physics, 1098 XG Amsterdam, The Netherlands}

\author{Maarten C. Mooij}
\affiliation{Nikhef, National Institute for Subatomic Physics, 1098 XG Amsterdam, The Netherlands}
\affiliation{Department of Physics and Astronomy, VU University Amsterdam, 1081 HV Amsterdam, The Netherlands}

\author{Yuly Chamorro}
\affiliation{Van Swinderen Institute for Particle Physics and Gravity, University of Groningen, 9747 AG Groningen, The Netherlands}

\author{Parul Aggarwal}
\altaffiliation[Current address: ]{JILA, National Institute of Standards and Technology and
University of Colorado, Boulder, CO 80309, USA}
\affiliation{Van Swinderen Institute for Particle Physics and Gravity, University of Groningen, 9747 AG Groningen, The Netherlands}
\affiliation{Nikhef, National Institute for Subatomic Physics, 1098 XG Amsterdam, The Netherlands}

\author{Hendrick L. Bethlem} 
\affiliation{Van Swinderen Institute for Particle Physics and Gravity, University of Groningen, 9747 AG Groningen, The Netherlands}
\affiliation{Department of Physics and Astronomy, VU University Amsterdam, 1081 HV Amsterdam, The Netherlands}

\author{Alexander Boeschoten}
\affiliation{Van Swinderen Institute for Particle Physics and Gravity, University of Groningen, 9747 AG Groningen, The Netherlands}
\affiliation{Nikhef, National Institute for Subatomic Physics, 1098 XG Amsterdam, The Netherlands}

\author{Anastasia Borschevsky}
\email{a.borschevsky@rug.nl}
\affiliation{Van Swinderen Institute for Particle Physics and Gravity, University of Groningen, 9747 AG Groningen, The Netherlands}
\affiliation{Nikhef, National Institute for Subatomic Physics, 1098 XG Amsterdam, The Netherlands}

\author{Kevin Esajas}
\affiliation{Van Swinderen Institute for Particle Physics and Gravity, University of Groningen, 9747 AG Groningen, The Netherlands}
\affiliation{Nikhef, National Institute for Subatomic Physics, 1098 XG Amsterdam, The Netherlands}

\author{Yongliang Hao}
\altaffiliation[Current address: ]{School of Physics and Electronic Engineering, Jiangsu University, Zhenjiang 212013, Jiangsu, China}
\affiliation{Van Swinderen Institute for Particle Physics and Gravity, University of Groningen, 9747 AG Groningen, The Netherlands}
\affiliation{Nikhef, National Institute for Subatomic Physics, 1098 XG Amsterdam, The Netherlands}

\author{Steven Hoekstra}
\affiliation{Van Swinderen Institute for Particle Physics and Gravity, University of Groningen, 9747 AG Groningen, The Netherlands}
\affiliation{Nikhef, National Institute for Subatomic Physics, 1098 XG Amsterdam, The Netherlands}

\author{Joost W. F. van Hofslot}
\affiliation{Van Swinderen Institute for Particle Physics and Gravity, University of Groningen, 9747 AG Groningen, The Netherlands}
\affiliation{Nikhef, National Institute for Subatomic Physics, 1098 XG Amsterdam, The Netherlands}

\author{Virginia R. Marshall}
\affiliation{Van Swinderen Institute for Particle Physics and Gravity, University of Groningen, 9747 AG Groningen, The Netherlands}
\affiliation{Nikhef, National Institute for Subatomic Physics, 1098 XG Amsterdam, The Netherlands}

\author{Thomas B. Meijknecht}
\affiliation{Van Swinderen Institute for Particle Physics and Gravity, University of Groningen, 9747 AG Groningen, The Netherlands}
\affiliation{Nikhef, National Institute for Subatomic Physics, 1098 XG Amsterdam, The Netherlands}

\author{Rob G. E. Timmermans}
\affiliation{Van Swinderen Institute for Particle Physics and Gravity, University of Groningen, 9747 AG Groningen, The Netherlands}
\affiliation{Nikhef, National Institute for Subatomic Physics, 1098 XG Amsterdam, The Netherlands}

\author{Anno Touwen}
\affiliation{Van Swinderen Institute for Particle Physics and Gravity, University of Groningen, 9747 AG Groningen, The Netherlands}
\affiliation{Nikhef, National Institute for Subatomic Physics, 1098 XG Amsterdam, The Netherlands}

\author{Wim Ubachs}
\affiliation{Department of Physics and Astronomy, VU University Amsterdam, 1081 HV Amsterdam, The Netherlands}

\author{Lorenz Willmann}
\affiliation{Van Swinderen Institute for Particle Physics and Gravity, University of Groningen, 9747 AG Groningen, The Netherlands}
\affiliation{Nikhef, National Institute for Subatomic Physics, 1098 XG Amsterdam, The Netherlands}

\author{Yanning Yin}
\altaffiliation[Current address: ]{Department of Chemistry, University of Basel, Klingelbergstrasse 80, 4056, Basel, Switzerland}
\affiliation{Van Swinderen Institute for Particle Physics and Gravity, University of Groningen, 9747 AG Groningen, The Netherlands}
\affiliation{Nikhef, National Institute for Subatomic Physics, 1098 XG Amsterdam, The Netherlands}

\collaboration{NL-\textit{e}EDM collaboration}

\begin{abstract}
We present an investigation of the performance of the relativistic multi-reference Fock-space coupled cluster (FSCC) method for predicting molecular hyperfine structure (HFS) constants, including a thorough computational study to estimate the associated uncertainties. In particular, we considered the $^{19}$F HFS constant in the ground and excited states of BaF. Due to a larger basis set dependence, the uncertainties on the excited state results (16-85\%) were found to be significantly larger than those on the ground state constants ($\sim$2\%).

The \textit{ab initio} values were compared to the recent experimental results, and good overall agreement within the theoretical uncertainties was found.  
This work demonstrates the predictive power of the FSCC method and the reliability of the established uncertainty estimates, which can be crucial in cases where the calculated property cannot be directly compared to experiment. 
\end{abstract}

\maketitle

\section{Introduction}
Accurate knowledge of excited state properties of atoms, ions, and molecules is necessary in many areas of precision physics for development of efficient laser cooling schemes \cite{Fitch2021}, construction of ever more precise atomic clocks \cite{Ludlow2015}, investigations of highly charged ions \cite{Kozlov2018}, and symmetry violating studies in molecules \cite{SafBudDem18}. %in many other aspects. 
In this context, the use of sophisticated \textit{ab initio} theoretical methods is crucial for providing accurate and reliable predictions of various parameters where experiment is not yet available, for interpretation of measurements, and for development of novel and highly sensitive measurement schemes.
This work is concerned with the reliability of \textit{ab initio} methods for predicting excited state molecular properties.

The most widely theoretically studied excited state properties are the electronic transition energies and the accompanying transition dipole moments, which are needed for the prediction and the interpretation of a wide range of spectra. These can be calculated with most \textit{ab initio} methods and to great precision. 
On the other hand, since accurate calculations on excited state magnetic hyperfine structure (HFS) constants in molecules are less tractable and computationally more involved, they are much more scarce to be found in literature compared to those in atoms.

We have previously carried out a theoretical investigation of the ground state hyperfine structure constants in BaF~\cite{Haase2020} using the relativistic coupled cluster (RCC) approach. The focus of that work was to benchmark the performance of RCC for these properties and to evaluate the reliability of the uncertainties that we assigned on our theoretical predictions using an extensive computational study. For the ground state HFS constants, we found our results to be in excellent agreement (well within the assigned theoretical uncertainties) with the available experimental values.

Recently, the HFS in the excited $^2\Pi$ state manifold of BaF was measured (Ref. \cite{Ginny}) as part of a preparatory study for an EDM measurement by the NL-\textit{e}EDM collaboration \cite{Aggarwal2018}. These measurements are particularly interesting as they allow us to extend our study to benchmarking the performance of the RCC approach for excited state HFS constants. It is expected that the excited state HFS constants will exhibit a different dependence on the computational parameters compared to the ground state parameters, due to the different spatial extent of the excited and ground state wave functions. 

Such benchmarks of \textit{ab initio} methods, carried out on measurable properties, also serve to confirm the accuracy of the approach for properties where no experiment is yet possible. For example, in case of BaF, such properties are the $P,T$-odd molecular enhancement factors, $W_s$ and $W_d$~\cite{Haase2021}, which are crucial for the interpretation of the NL-eEDM experiment \cite{Aggarwal2018}, or the $W_A$ factor \cite{HaoIliEli18}, needed for extracting the nuclear anapole moment from future measurements~\cite{AltAmmCah18}.
BaF is particularly interesting for precision experiments due to, among other things, its good laser cooling properties \cite{Chen2016,Albrecht2020,Kogel2021}.

Reliable \textit{ab initio} methods for calculating excited state HFS constants should treat electron correlation on a high level, should employ a large, sufficiently converged basis set, and, for heavy-atom containing systems like BaF, should also be carried out in a relativistic framework. For atoms, several such implementations exist, based on the relativistic configuration interaction approach (CI) (with or without the many-body perturbation theory corrections (+MBPT))~\cite{Dzuba1996,Dzuba1998}, the relativistic Hartree-Fock method with MBPT~\cite{Dzuba1984,Dzuba1987,Ginges2017}, the various variants of the relativistic coupled cluster approach~\cite{Safronova1998,Safronova1999,Das2011}, or the multiconfiguration Dirac-Fock method~\cite{Bieron2004,Jonsson2007}. For molecules however, such methods are limited. To the best of our knowledge, the only recent example of accurate HFS calculations in molecules is the work of Oleynichenko et al.~\cite{Oleynichenko2020}, which presents relativistic Fock-space coupled cluster (FSCC) calculations of the HFS constants in the excited states of the KCs molecule.

A challenging aspect of theoretical \textit{ab initio} investigations is a lack of a straightforward scheme for performing a reliable uncertainty evaluation, which is consequently absent in many studies. However, for comparison of the calculated properties with experiments and with other theoretical methods, such uncertainty estimate is necessary and it should preferably be obtained from purely theoretical considerations. Only a method that is transparent (in that it is clear which approximations are made and which effects are included) and that can be systematically improved allows for a reliable theoretical uncertainty evaluation.

In this study, we investigate the performance of the relativistic FSCC method in calculating excited state HFS constants and estimating the theoretical uncertainty. The FSCC method has shown excellent performance in predicting energies and spectroscopic constants of molecules \cite{Lyakh2012, Hao2018, Hao2019, Oleynichenko2020, Pototschnig2021a} and has recently been applied by us to the  HFS constants of excited states in atoms by making use of the finite-field scheme \cite{Gustafsson2020, Kanellakopoulos2020, DeVries}. 
In addition, we use the single-reference coupled-cluster method with single, double, and perturbative triple excitations (CCSD(T)) to calculate the ground state HFS constants. This method has recently shown excellent agreement (below 1\%, with a conservative uncertainty estimate of 5.5\%) between calculated and experimental HFS constants for the ground states of $^{133}$Cs and $^{137}$BaF (i.e. coupling to the $^{137}$Ba nucleus in BaF) \cite{Haase2020}. 

Here, we focus on the $^{138}$Ba$^{19}$F HFS constants in the $X^2\Sigma^+$ ground state and the $A^2\Pi_{1/2}$ and $A^2\Pi_{3/2}$ excited states, motivated by the recently performed measurements \cite{Ginny}. Compared to Ref. \citenum{Haase2020}, the present study puts higher requirements on the theoretical method for two reasons: 1) As we are dealing with the hyperfine structure due to the coupling to the fluorine, we need to provide a high-quality description both of the surroundings of its nucleus, and of the bond area; 2) For the excited state HFS constants, the FSCC method should be used and an even stronger dependence on the quality of the basis set can be expected.

We present a scheme for estimating the theoretical uncertainty that aims to quantify the approximations of the employed method through a careful analysis of the effect of the different computational parameters on the calculated HFS constants.

In the last part, we derive relations between the \textit{ab initio} HFS constants and the Frosch \& Foley parameters, frequently used in molecular spectroscopy \cite{Frosch1952}
and finally, we  compare the \textit{ab initio} results to the experimental HFS constants,  obtained from an analysis of the measured energy splittings from Ref.~\cite{Ginny}.

\section{Theory}\label{sec:BaF_HFS_theory}

The finite-field relativistic coupled-cluster method for calculation of magnetic HFS parameters is described in detail in Ref.~\cite{Haase2020} and we will briefly recap the methodology here. Throughout this section atomic units are used.

In axial systems (with the symmetry axis along the $z$-axis) the HFS tensor due to nucleus $A$ can be parametrized in terms of the parallel,
\begin{equation}\label{eq:Apar}
A_{\parallel}=\frac{\mu_A}{I_A\tilde{S}_z}\left\langle
\sum_i \frac{( \vec{r}_{iA} \times \vec{{\alpha}_i} )_z}{r_{iA}^3}
\right\rangle_{\psi^{(z)}},
\end{equation}
and the perpendicular component,
\begin{equation}\label{eq:Aperp}    
A_{\perp}=\frac{\mu_A}{I_A\tilde{S}_{x/y}}\left\langle
\sum_i \frac{( \vec{r}_{iA} \times \vec{{\alpha}_i} )_{x/y}}{r_{iA}^3}
\right\rangle_{\psi^{(x/y)}},
\end{equation}
where $\mu_A$ and $I_A$ are the nuclear magnetic moment (in units of the nuclear magneton $\mu_N=(2m_pc)^{-1}$) and the nuclear spin, respectively, of nucleus $A$. 
$\tilde{S}_z$ and $\tilde{S}_{x/y}$ are the projections of the total electronic angular momentum onto the $z$- or $x/y$- axes.
$\vec{\alpha}_i$ is the Dirac matrix defined below and $\vec{r}_{iA}$ is the vector between electron $i$ and nucleus $A$. The superscript on the electronic wave function, $\psi^{(u)}$, indicates the quantization axis of the total electronic angular momentum. The HFS parameters, $\Apar$ and $\Aperp$, can further be defined by an effective Hamiltonian, which will be introduced in Eq. (\ref{eq:AparAperp}).

In this work, the electronic wave function is calculated using the single reference as well as the Fock-space coupled-cluster methods. In the case of coupled-cluster wave functions, the calculation of expectation values is not straightforward. Instead, we evaluate these constants by means of the finite-field method in which the HFS Hamiltonian,
\begin{equation}
H^{\text{HFS}}_u =
\frac{\mu_A}{I_A\Omega} \sum_i \frac{( \vec{r}_{iA} \times \vec{{\alpha}_i} )_{u}}{r_{iA}^3} ,
\end{equation}
with $u$ corresponding to $z$ in the case of $\Apar$ and $x$ or $y$ in the case of $\Aperp$, see Eq. (\ref{eq:Apar}) and (\ref{eq:Aperp}) respectively, is added to the unperturbed Dirac-Coulomb Hamiltonian: 
\begin{equation}
H^{(0)}=\sum_i\left[c\vec{\alpha}_i\cdot\vec{p}_i+\beta_i c^2+V_{\text{nuc}}(\vec{r}_i)\right]
+ \frac{1}{2} \sum_{i\neq j}\frac{1}{r_{ij}},
\end{equation}
with a prefactor $\lambda$, referred to as the field strength:
\begin{equation}
H = H^{(0)} + \lambda_u H_u^{\text{HFS}} .
\end{equation}
Here $\vec{\alpha}=\begin{pmatrix} 0_{2\times2} & \vec{\sigma} \\ \vec{\sigma} & 0_{2\times2} \end{pmatrix}$ and $\beta=\begin{pmatrix} 1_{2\times2} & 0_{2\times2} \\ 0_{2\times2} & -1_{2\times2} \end{pmatrix}$ are the usual Dirac matrices, $\vec{\sigma}$ the vector of Pauli spin matrices and $V_{\text{nuc}}(\vec{r}_{i})$ is the nuclear potential in the form of a Gaussian charge distribution \cite{Visscher1997b} and $\vec{r}_i$ is the position vector of electron $i$.

The values of $\lambda$ should be small enough to prevent higher order terms contributing to the energy and large enough to avoid numerical issues. Consequently, the HFS parameters can be obtained by the derivative of the total energy with respect to $\lambda$:  
\begin{equation}
\label{hyp_ff}
A_{\parallel(\perp)}= \left.\frac{dE^{z(x/y)}(\lambda_{z(x/y)})}{d \lambda_{z(x/y)}}\right|_{\lambda_{z(x/y)}=0}.
\end{equation}
Within the finite-field method this derivative is evaluated numerically.

\section{Computational details}
The \textit{ab initio} computational study was performed using the DIRAC17 program package \cite{Dirac17}. The ground state experimental equilibrium bond length $d_\text{Ba-F}=2.162$\AA{} \cite{Knight1971} was employed for both the ground and excited states. The magnetic moment of fluorine was taken as $\mu(^{19}\text{F})=2.628868\mu_N$ \cite{Stone2005}.

The magnetic hyperfine interaction constants $A_{\parallel}$ and $A_\perp$ were calculated by employing the finite-field approach, introduced in the previous section.
The same calculation was repeated three times, with field strengths ($\lambda$) of $-10^{-5}$, $0$, and $10^{-5}$ for the perpendicular component and $-10^{-4}$, $0$, and $10^{-4}$ for the parallel component. These field strengths were chosen to ensure a linear behavior of the total energy with respect to the perturbation.
The convergence criterion of the coupled cluster amplitudes was set to $10^{-12}$ a.u.
The HFS constants were obtained from the derivative of the energy with respect to the field strength by the means of linear regression through the three points.

Electron correlation was treated within the different variants of the coupled cluster approach, i.e. the relativistic CCSD(T) and FSCC. 
In the case of the single-reference CC method, three different schemes were used for including the perturbative triple excitations: The CCSD(T) method~\cite{Raghavachari1989}) includes all fourth-order terms and part of the fifth-order terms, while the CCSD-T method~\cite{Deegan1994} includes one additional fifth-order term. The CCSD+T method~\cite{Urban1985} includes fourth-order terms only.
In the coupled cluster calculations all electrons were included in the correlation treatment and the virtual space cutoff was set to 2000 a.u., if not stated otherwise.

 Dyall's relativistic uncontracted basis sets of varying quality (double to quadruple-zeta) \cite{Dyall2009,Gomes2010,Dyall2016} were employed in the calculations, with or without additional core-correlating functions (cv$n$z vs v$n$z). We also tested the effect of augmentation of these standard basis sets by adding, in an even-tempered fashion,  further diffuse (low exponent) and tight (high exponent) basis functions.

\section{Results}

In the following sections we present a computational study of the basis set and electron correlation effects. The results of these investigations will be used in Sec. \ref{sec:uncertain} to determine the most suitable computational scheme and to establish a reliable uncertainty estimate. 

In all calculations, the perpendicular component, $\Aperp$, in the $^2\Pi_{3/2}$ state was zero.
%and we will consequently omit it in the presented results. 
As will be discussed in Sec. \ref{sec:FenF}, the vanishing value for $\Aperp$ in this state is related to the vanishing $\Lambda$-doubling.

 \subsection{Basis sets}
In order to obtain the $^{19}$F HFS constants in the ground and excited state of BaF with high accuracy, the chosen basis set has to fulfill many requirements. 
The unpaired electron occupies a molecular orbital of mainly Ba character, which couples to the ligand F atom. Consequently, both the valence orbital, the bonding area, and the area of the F nucleus should be sufficiently covered by the basis functions. In addition, when considering the excited $\Pi$ states, the wave function can be expected to expand and an even larger spatial area should be taken into account. In the following, we will investigate the performance of the Dyall basis sets with and without additional functions to properly describe this challenging situation. The results presented in this section were obtained with FSCC (0,1) method, since this method is applicable to both the ground and the excited states, as will be discussed in the following section. 

\begin{figure}[t]
    \centering
    \includegraphics[width=\columnwidth]{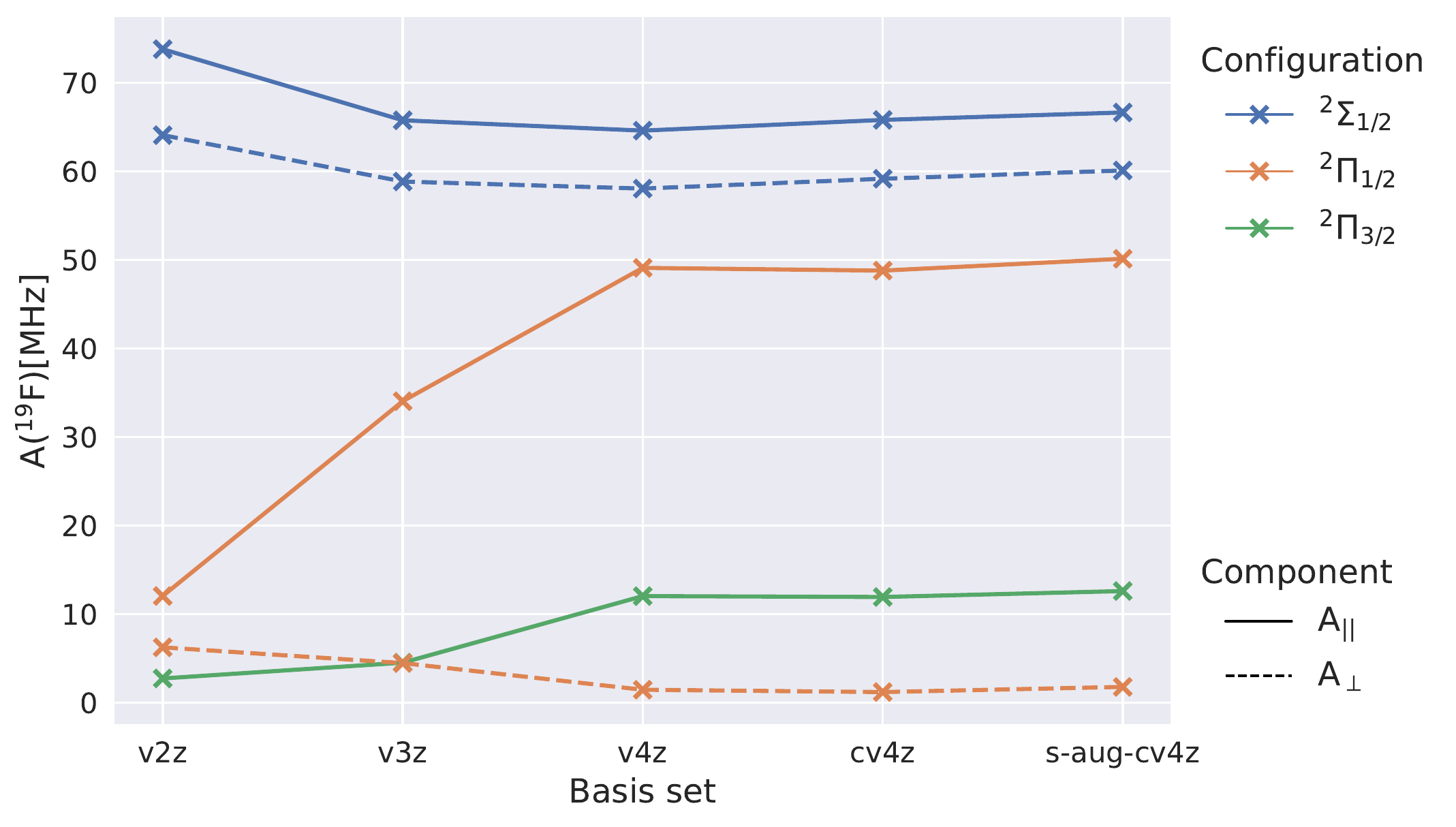}
    \caption{Effect of the basis quality on the calculated magnetic hyperfine interaction constants of the F nucleus [MHz] in the three lowest-lying states of BaF.  The FSCC(0,1) method was used with all electrons correlated and a virtual cutoff of 2000 a.u. 
    }
    \label{fig:Aparbasis}
\end{figure}

In Table~\ref{tab:Abasissize} and Fig.~\ref{fig:Aparbasis}, the effect of increasing the size (i.e. the cardinality) of the basis set on the calculated HFS constants is shown. The first thing to notice is that, as expected, the basis set dependence is much larger for the excited states compared to the ground state. 
In the case of the ground state, both components, $\Apar$ and $\Aperp$, decrease with increasing basis set size. For the excited states $\Apar$ increases and $\Aperp$ decreases with increasing basis set size.

\begin{table}[t]
\renewcommand*{\arraystretch}{1.25}
    \centering
    \caption{Calculated magnetic hyperfine interaction constants of the F nucleus [MHz] in the three lowest-lying states of BaF for increasing quality
basis sets. The FSCC(0,1) method was used with all electrons correlated and a virtual cutoff of 2000 a.u. } \label{tab:Abasissize}
    \begin{tabular}{lccccc}
    \hline
  & \multicolumn{2}{c}{$^2\Sigma_{1/2}$} & \multicolumn{2}{c}{$^2\Pi_{1/2}$} & {$^2\Pi_{3/2}$} \\
    \hline
  &  A$_{\parallel}$  & A$_{\perp}$  &  A$_{\parallel}$  & A$_{\perp}$  &  A$_{\parallel}$  \\
    \cline{2-6}
  v2z~ &73.80& 64.09 &12.04&6.24 &2.73	 \\
  v3z  &65.77&58.85	& 34.04 & 4.47& 	4.51  \\
  v4z  & 64.60 &	58.05 &49.10 & 1.45	&12.03 \\
    \hline
    \end{tabular}
\end{table}

For both parallel and perpendicular components, the $^2\Sigma_{1/2}$ ground state values show a converging behavior with increasing basis set size. Going from v2z to v3z lowers the values by $12\%$ and $9\%$ for A$_{\parallel}$ and A$_{\perp}$, respectively, while going to v4z changes the values by a much smaller amount, namely $1.8\%$ and $1.4\%$. The situation is very different for the excited states for which the HFS parameters do not seem to be converged at the v4z level, as is clear in Fig. \ref{fig:Aparbasis}. 

The completeness of the v4z basis set was checked by adding extra functions that improve the description of the electronic wave function in the vicinity of the nucleus. In order to do so, we used the core-valence basis set (cv4z) where six Ba (3f,2g,1h) and three F (2d,1f) functions were added to the v4z basis set, as well as the all-electron (ae4z) basis set where nine additional Ba functions (5f,3g,1h) were added. 
In a recent study on the sensitivity to the parity violating nuclear anapole moment in light elements \cite{HaoNavNor20}, the effect of adding tight s- and p-functions on the F-atom, was found to be important. By contrast, in this study the effect of adding these functions resulted in a minor change of less than $\sim$0.1\%.
Furthermore, the quality of the description in the valence region was tested by using the augmented basis sets (s-aug-v4z, d-aug-v4z and t-aug-v4z) in which one, two or three diffuse functions in each symmetry were added in an even-tempered fashion. 
The calculations of $\Aperp$ demand considerably more computational resources, due to the required lower symmetry for the this property, compared to the calculations of $\Apar$. 
Consequently, the calculations using the largest basis sets, i.e. ae4z, d-aug-v4z and t-aug-v4z, were performed only for $\Apar$ and it is assumed that the observed effects hold for $\Aperp$ as well. 
The results are displayed in Table~\ref{tab:Abasisaug}.  

In the ground state, the addition of one set of diffuse functions and the change to the cv4z basis set result in small corrections of 1.4\% and 1.7\%, respectively, in the case of $\Apar$. Similar values are obtained for $\Aperp$, which justifies our assumption that the effect of further augmentations on this component would be of similar magnitude as on $\Apar$. 
In the case of the cv4z and s-aug-v4z basis sets, we investigated the effects of adding extra functions on the individual atoms. Again, due to computational costs, we considered only $\Apar$. These results are also presented in Table~\ref{tab:Abasisaug}.
In the case of diffuse functions, which improve the description of the bonding region, we find that mainly those on the F atom have a discernible effect on the calculated HFS constants. Addition of further sets of diffuse functions has a negligible effect on the ground state properties.

For the excited states the effect of adding diffuse functions is larger, especially for $\Aperp$, namely 2.7\% and 41\% for $\Apar$ and $\Aperp$ in the $^2\Pi_{1/2}$ state and 5.6\% for $\Apar$ in $^2\Pi_{1/2}$ state. Contrary to the ground state, this effect is mainly due to the diffuse functions of Ba. The addition of a second set of diffuse functions results in a further increase of 1.9\% and 2.4\% for $\Apar$ in the $^2\Pi_{1/2}$ and the $^2\Pi_{3/2}$ states and the effect of the addition of a third set is less than 0.4\%. For the excited states the addition of diffuse functions is consequently reasonably converged at the d-aug-v4z level. However, in view of the additional computational costs, we choose to present the recommended values with one additional set of diffuse (s-aug-) functions and we account for the missing effects in the uncertainty estimate. 

Switching to the cv4z basis set results in an effect of -0.63\% and -18\% for $\Apar$ and $\Aperp$ in the $^2\Pi_{1/2}$ state and of -0.83\% for $\Apar$ in the $^2\Pi_{1/2}$ state. As for the ground state, the effect is dominated by the additional functions on Ba. Interestingly, going to the ae4z basis set, a similar but opposite effect of 0.92\% and 1.0\% for $\Apar$ in the $^2\Pi_{1/2}$ and $^2\Pi_{3/2}$ states is observed, leaving the v4z and the ae4z results almost identical. Again, in view of computational costs, we choose to employ the cv4z basis sets for the final recommended values.  

To summarize, we observed a larger basis set dependence of the calculated excited state HFS constants compared to the ground state, which is in line with our expectations. Balancing the level of accuracy and the computational costs, we choose to use the s-aug-cv4z for the recommended values. For this basis set diffuse functions of high angular momentum (in this case h-functions) were omitted in order to reduce computational costs.

\begin{table}[]
    \caption{Calculated magnetic hyperfine interaction constants of the F nucleus [MHz] in the three lowest-lying states of BaF for various augmentations of the v4z basis sets. For the cv4z and s-aug-v4z basis set, the addition of extra functions were investigated for the two atoms individually.  The FSCC(0,1) method was used with all electrons correlated and a virtual cutoff of 2000 a.u.}
    \label{tab:Abasisaug}
    \centering
        \begin{tabular}{lccccc}
    \hline
          & \multicolumn{2}{c}{$^2\Sigma_{1/2}$} & \multicolumn{2}{c}{$^2\Pi_{1/2}$} & {$^2\Pi_{3/2}$} \\
  \hline
  &  A$_{\parallel}$  & A$_{\perp}$  &  A$_{\parallel}$  & A$_{\perp}$  &  A$_{\parallel}$    \\
  \cline{2-6}
  v4z       & 64.60& 58.05	&49.10 & 1.45	&12.03 \\
  ~ Ba cv4z F v4z      & +1.23 &  & -0.37 &  & -0.08\\
  ~ Ba v4z F cv4z      & -0.03 &  & +0.07 &  & -0.02\\
  cv4z      &65.81& 59.17 &48.79& 1.19 &11.93\\
  ae4z      & 65.71 & & 49.24 & &12.05 \\
  ~ Ba s-aug-v4z F v4z &+0.01  &  &+1.10  &  &+0.51\\
  ~ Ba v4z F s-aug-v4z &+0.86  &  &+0.32  &  &+0.18\\
  s-aug-v4z &65.44 & 59.00	&50.45 & 2.05 &12.71 \\
  d-aug-v4z &65.52 &	&51.42 &	&13.02\\
  t-aug-v4z & 65.51&	& 51.57 &	&13.07 \\
s-aug-cv4z& 66.65 & 60.10 & 50.13 & 1.78 & 12.60 \\
         \hline
    \end{tabular}
\end{table}

\subsection{Electron correlation}\label{sec:corr}

We use the optimal basis set, determined above, to test  the effect of treatment of correlation on the calculated HFS constants. For the ground $^2\Sigma_{1/2}$ state, the single-reference CCSD method with the three different schemes for including perturbative triple excitations, CCSD+T, CCSD(T) and CCSD-T, can be used. Application of this method to the excited states resulted in large T1-values \cite{Lee1989} ($\sim$ 0.16) indicating the need for a multi-reference method. In this work, we used the multi-reference FSCC approach (sector (0,1)) for which the starting point is a CCSD calculation on the BaF$^{+}$ ion after which a single electron is added to the orbitals in the so-called model space, here consisting of the three lowest spinors, i.e. $\sigma$, $\pi_{1/2}$ and $\pi_{3/2}$.

 \begin{table}[t]
     \caption{
     Calculated magnetic hyperfine interaction constants of the F nucleus [MHz] in the ground state of BaF obtained with the s-aug-cv4z basis set, all electrons correlated, a 2000 a.u. virtual space cutoff, and different correlation methods.}
     \label{tab:Acorr}
     \centering
         \begin{tabular}{lcc}
     \hline
           & \multicolumn{2}{c}{$^2\Sigma_{1/2}$} \\ 
     \hline
   &  A$_{\parallel}$  & A$_{\perp}$ \\ 
   \cline{2-3}
   CCSD & 67.11 &  60.87     \\ 
   CCSD(T) & 71.22 & 64.02   \\ 
   CCSD+T & 70.90 & 64.06    \\ 
   CCSD-T & 71.22 & 63.62    \\ 
   FSCC(0,1) & 66.65 & 60.10 \\ 
          \hline
     \end{tabular}
 \end{table}

Table~\ref{tab:Acorr} contains the comparison of the different single reference CC schemes and the FSCC (0,1) results for the ground $^2\Sigma_{1/2}$ state.
The values obtained with the CCSD and the FSCC (0,1) methods are in a very good agreement with each other, with a difference of 0.7\% and 1.3\% for $\Apar$ and $\Aperp$, respectively, which indicates the strong single-reference nature of this state and justifies the use of the single-reference method. 
The inclusion of perturbative triples has a significant effect on the calculated values. Compared to the CCSD results, the use of the CCSD(T) method leads to an increase of 6\% for $\Apar$ and 5\% for $\Aperp$.
The spread in the results obtained with the three different schemes for perturbative triples is however small, i.e. within 0.4\% for $\Apar$ and 0.7\% for $\Aperp$. This is an indication that the effect of higher order excitations is small. 

It is interesting to compare the correlation effects seen here with the ones presented in our previous study on the $^{137}$Ba HFS constant in BaF \cite{Haase2020}, since the two cases show very different behavior. In the case of the $^{137}$Ba coupling, the spread in the results obtained with the different perturbative triples methods was larger than the effect of including pertubative triples in the first place. We consequently concluded that the inclusion of perturbative triples lead to unstable results and we chose the CCSD method for the recommended values. Here, in contrast, the negligible difference in the perturbative triple excitation contribution from the different schemes gives us confidence in the CCSD(T) approach. 

In the results presented thus far, all electrons were correlated and consequently a virtual cut-off of 2000 a.u. was used. We investigated the effect of including even higher lying virtual orbitals by increasing this cut-off to 10.000 a.u. (for $\Apar$ at the v3z level). For all the states considered, the observed effect was not larger than 0.03\% meaning that a cut-off of 2000 a.u. is indeed suitable.

\subsection{Recommended values and uncertainty estimation} \label{sec:uncertain}

Based on the computational study presented in the previous sections, we use the FSCC (0,1) approach for the $\Pi$ states, while the CCSD(T) method is most suitable for the $\Sigma$ ground state; the final recommended values are obtained using the s-aug-cv4z basis set.  
For both approaches, all electrons were correlated and a virtual space cut-off of 2000 a.u. was used. The recommended values are presented in Table \ref{tabfinal} along with theoretical uncertainty estimates, discussed in the next section. %In the following we will elaborate on the scheme for determining these uncertainties and in 
Section~\ref{sec:comparison} contains the comparison of the calculated HFS constants with the experimental values.

\begin{table}[t]
 \centering
 \caption{Description of the scheme employed to estimate the uncertainty due to each computational parameter. }
 \label{taberrormethod}
 \begin{tabular}{lll}
 \hline
{Error source}	    	& \multicolumn{2}{l}{Description}	\\
\hline
   \multicolumn{2}{l}{\textbf{Basis set}}	\\
     ~~Quality & & $\frac{1}{2}$ $\cdot$ (v4z-v3z) \\ 
     ~~Core-corr.  & A$_{\parallel}$: & ae4z-cv4z \\ \noalign{\smallskip}
     ~~functs. & A$_{\perp}$: & $\frac{\text{ae4z}_{\text{A}_{\parallel}} - \text{cv4z}_{\text{A}_{\parallel}}}{\text{cv4z}_{\text{A}_{\parallel}}}$ 
    $\cdot$ cv4z$_{\text{A}_{\perp}}$  \\ 
    ~~Diffuse functs. & A$_{\parallel}$: & t-aug-v4z - s-aug-v4z \\ \noalign{\smallskip}
    & A$_{\perp}$: & $\frac{\text{ t-aug-v4z}_{\text{A}_{\parallel}} - \text{s-aug-v4z}_{\text{A}_{\parallel}}}{\text{s-aug-v4z}_{\text{A}_{\parallel}}}$     $\cdot$ s-aug-v4z$_{\text{A}_{\perp}}$ \\ 
\multicolumn{2}{l}{\textbf{Correlation}} \\
 ~~Virt. cut-off &  & 10.000 au. - 2000 au. (v3z) \\ 
 ~~Triples & $^2\Sigma_{1/2}$:~ & $\frac{1}{2}$ $\cdot$ (CCSD(T) - CCSD)  \\ \noalign{\smallskip}
            & $^2\Pi$: & 2 $\cdot$ $\frac{\text{CCSD(T)}_{\Sigma} - \text{CCSD}_{\Sigma}}{\text{FSCC}_{\Sigma}}$
    $\cdot$ FSCC$_{\Pi}$  \\ 
\noalign{\smallskip}
\hline
 \end{tabular} 
\end{table}

\subsubsection{Strategy for estimating uncertainties}

The main sources of uncertainty in our calculations are the incompleteness of the basis set, the neglect of higher excitations (beyond doubles for FSCC and beyond (T) for CCSD(T)), the virtual space cutoff in the correlation treatment, and the incomplete treatment of relativity. The uncertainty due to the various approximations can be separately estimated for each computational parameter as the difference between the result obtained using the best feasible model and the second best. A similar scheme was used in our earlier works on symmetry breaking properties~\cite{Hao2018, Denis2019, Haase2020, Denis2020}. The specific schemes for estimating the uncertainties for each computational parameter are summarised in Table~\ref{taberrormethod} and elaborated in the following. 

\paragraph{Basis set}

The uncertainty due to the choice of a finite basis set can be divided into three contributions: the quality (indicated by the cardinal number), the amount of core-correlating functions and the amount of diffuse functions. 

In order to estimate the missing effects due to using a 4z level basis set the difference between the v4z and v3z results is considered. Since we expect a converging behavior when increasing the basis set size, we take half of this difference as the corresponding uncertainty. 

To estimate the uncertainty related to the additional functions, we consider the difference between the ae4z and cv4z basis sets for the core-correlating functions and between t-aug-v4z and s-aug-v4z basis sets for the diffuse functions.  

Due to the high computational costs, $\Aperp$ was not calculated using the largest ae4z and t-aug-v4z basis sets. We consequently work on the assumption that the two components, $\Apar$ and $\Aperp$, experience a similar effect due to the addition of extra functions and the uncertainties for $\Aperp$ were determined by using the relative uncertainty for $\Apar$ (see Table IV). 

\paragraph{Virtual space cut-off}

To estimate the uncertainty due to the active space cutoff applied in our calculations, we take the difference between the values obtained with a cutoff of $10000$ a.u. and $2000$ a.u. at the FSCC v3z level, assuming that further increase of the active space (i.e. including all the virtual orbitals in the correlation space) should have a significantly smaller effect on the results.

\paragraph{Higher-order excitations}

In the $^2\Sigma_{1/2}$ ground state the error due to the missing triple and higher excitation is estimated by taking half the difference between CCSD(T) and CCSD, similar to our previous works~\cite{Hao2018, Denis2019, Haase2020, Denis2020,Haase2021}. The factor 1/2 is justified by the small spread in the results obtained with different treatment of perturbative triples.

This procedure cannot be followed for the $\Pi$ states investigated here (as the current FSCC implementation includes single and double excitations only) and an alternative approach was adopted for these states.
We estimate this uncertainty based on the relative effect observed in the ground state. Since correlation effects can be expected to be larger in the excited states we use twice the relative difference between the $^2\Sigma_{1/2}$ CCSD(T) and CCSD results and multiply this by the $^2\Pi_{1/2}$ and $^2\Pi_{3/2}$ FSCC values.

\begin{table}[bt]
 \centering
 \caption{Summary of the most significant contributions to the calculated magnetic hyperfine interaction constants [MHz] of the $^{19}$F nucleus in the three lowest-lying states of BaF. 
}
 \label{taberror}
 \begin{tabular}{llccccc}
 \hline
 &	& \multicolumn{2}{c}{$^1\Sigma_{1/2}$} & 	\multicolumn{2}{c}{$^2\Pi_{1/2}$} & $^2\Pi_{3/2}$	\\
\hline
\multicolumn{2}{l}{\textbf{Error source}} &  $\delta$A$_{\parallel}$  & $\delta$A$_{\perp}$  &  $\delta$A$_{\parallel}$  & $\delta$A$_{\perp}$  &  $\delta$A$_{\parallel}$ \\
\cline{3-7}
   \multicolumn{2}{l}{Basis set}	\\
& {Quality} & -0.59 & -0.40 & 7.53 & -1.51 & 3.76 \\ 
& Tight functions & -0.1 & -0.09 & 0.45  & 0.01 & 0.12 \\ 
& Diffuse functions & 0.07 & 0.06 & 1.12 & 0.04 & 0.36 \\ 
\multicolumn{2}{l}{Correlation} & & &\\
& Virtual space cut-off & 0.02 & 0.02 & 0.00 & 0.00 & 0.00 \\ 
& Triples & 2.05   & 1.57 & 3.09  & 0.09 & 0.78   \\ \noalign{\smallskip}
\cline{3-7} \noalign{\smallskip}
\multicolumn{2}{l}{\textbf{Total~~~}} & 2.14  & 1.62 & 8.23  & 1.51 & 3.86  \\ 
\multicolumn{2}{l}{\textbf{\%~~~}} & 3.00  & 2.54 & 16.41  & 85.16 & 30.62   \\ 
\hline
 \end{tabular}
\end{table}

\paragraph{Higher-order relativistic effects}

In our calculations, the electron-electron interaction is described using the non-relativistic Coulomb operator. Within the framework of the DIRAC program, a rough estimate of the Breit contribution can be extracted by including the Gaunt interaction in the calculation~\cite{Pernpointner2002}. However, this contribution can presently be calculated only on the Hartree-Fock level, which is not applicable to the excited states considered here, and consequently, we will omit this source of uncertainty. 

\subsubsection{Total theoretical uncertainties}

\begin{figure}[t]
    \centering
    \includegraphics[width=\columnwidth]{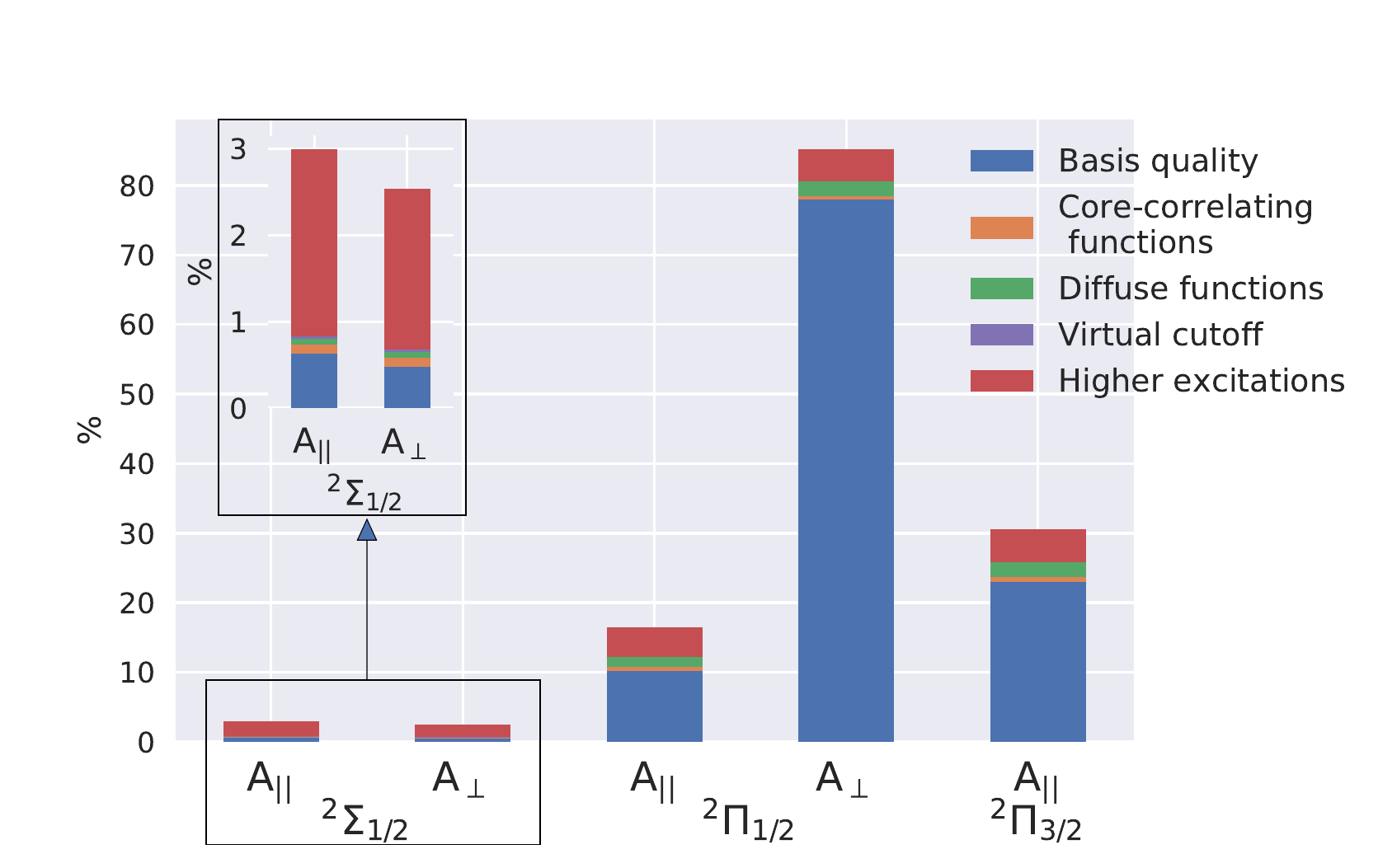}
    \caption{Contribution of the different sources of uncertainty to the total error bar in HFS constant, given in percentage of the final values.}
    \label{figerror}
\end{figure}

The individual contributions to the uncertainties, obtained using the strategy outlined above, are presented in Table~\ref{taberror} and Fig.~\ref{figerror}. We expect the individual uncertainties to be independent of each other to a good approximation, and thus the total uncertainty is obtained by a quadratic sum of the individual uncertainties. This assumption can be expected to be valid when the basis set and correlation effects are both treated on a sufficient level, which is the case here.  

We observe that the magnitude of the estimated uncertainties is dramatically different between the ground and the excited states, demonstrating the complexity of predicting properties of molecular excited states and highlighting the importance of a separate uncertainty evaluation for different electronic states. This difference is mainly due to the large and dominating contribution of the uncertainty due to the basis set quality in the excited states, as indicated by the blue bars in Fig. \ref{figerror}. For these states, the results are not converged on the 4z level. For the ground state, on the other hand, the uncertainty stemming from the from higher order excitations (red bars) dominates. 

\section{Comparison of theoretical and experimental results} 

Within the field of experimental molecular physics, the HFS of diatomic molecules is often described in terms of the Frosch \& Foley (F\&F) parameters \cite{Frosch1952}. Consequently, we will derive relations between the \textit{ab initio} and F\&F parameters for comparison of theoretical and experimental values.
Most equations needed for this analysis can be found in the literature \cite{Sauer1999, Frosch1952, Dousmanis1955, Brown2003, Mawhorter2011}; here we present a consistent and easy-to-follow derivation. In addition, we note that, to the best of our knowledge, the equations in terms of the \textit{ab initio} constants for the $^2\Pi_{3/2}$ state cannot be found elsewhere. 

In order to compare the theoretical and experimental parameters for the $\Pi$ states, we first need to extract the experimental HFS parameters from the measured results.
These are presented in Ref. \cite{Ginny} in terms of 6 measured hyperfine energy splittings of in total 4 different rotational levels within the excited $^2\Pi$ state manifold, see Table~\ref{tab:energies}. 

In Sections~\ref{sec:FenF} and \ref{sec:ab_initio} the energy splittings of the hyperfine levels are  derived starting from the F\&F and \textit{ab initio} Hamiltonians, respectively. In Section~\ref{sec:relations}, these energy splittings will be used to derive relations between the two sets of parameters. In Section~\ref{sec:fit}, the experimental results will be fitted in order to extract the HFS constants and in Section~\ref{sec:comparison} the resulting experimental and theoretical HFS parameters will be compared.

\subsection{Energy splittings in terms of the Frosch \& Foley parameters}\label{sec:FenF}

To derive the energy splittings, we first recap the Hamiltonian, derived by Frosch \& Foley \cite{Frosch1952}:
\begin{eqnarray}
H^{\text{F\&F}} &=& a \vec{I}\cdot\vec{L} + (b+c) I_z S_z + b (I_xS_x + I_yS_y) \nonumber \\ 
&+& \frac{1}{2} d \left( e^{2i\phi} I^-S^- + e^{-2i\phi} I^+S^+ \right),  
\label{eq:full_hfs}
\end{eqnarray}
where the internuclear axis is aligned with the $z$-axis, $a$, $b$, $c$, and $d$ are the molecular HFS constants and $\vec{I}$, $\vec{L}$ and $\vec{S}$ are the vectors describing nuclear spin, orbital angular momentum and electron spin, respectively. The term involving $d$ in Eq.~(\ref{eq:full_hfs}) is dominated by $\Lambda$-doubling. 
In the following, we will derive specific Hamiltonians for the $^2\Sigma$, $^2\Pi_{1/2}$ and $^2\Pi_{3/2}$ states. 

Since the $^2\Sigma$ state has no electronic orbital angular momentum, the term involving $a$ in Eq. (\ref{eq:full_hfs}) vanishes and the $\Lambda$-doubling term involving $d$ has no first-order contribution \cite{Dousmanis1955}. 
Therefore, the Hamiltonian reduces to:
\begin{eqnarray}
H^{\text{F\&F}}_{\Sigma} &=& (b+c) I_z S_z + b (I_xS_x + I_yS_y).
\end{eqnarray}

For BaF in the $\Pi$ manifold, the spin-orbit coupling constant, $A$, is much larger than the rotational constant, $B$, ($A = 632.161(1)$ cm$^{-1}$ and $B = 0.2117889(27)$ cm$^{-1}~$\cite{Effantin1990}) and it can be described best in Hund's case a). In Hund's case a), the only well-defined component of $\vec{L}$ is the $z$-component, denoted $\Lambda$, as it is strongly coupled to the internuclear axis.
The expectation value of the first two terms in Eq. (\ref{eq:full_hfs}) can thus be simplified to $h=[\Lambda a + \Sigma (b+c)]I_z$, where $\Sigma$ is the projection of the electron spin onto the internuclear axis.

Within the $\Pi$ manifold, the operator $(I_xS_x + I_yS_y)$ couples the $\Pi_{1/2}$ and $\Pi_{3/2}$ states. The term involving this operator can be neglected due to the large spin-orbit splitting between these two states. Consequently, for the $^2\Pi_{1/2}$ state, the Hamiltonian becomes: 
\begin{eqnarray} \label{eq:FF_1/2}
H^{\text{F\&F}}_{^2\Pi_{1/2}} &=& h_{1/2} I_z + \frac{1}{2} d \left( e^{2i\phi} I^-S^- + e^{-2i\phi} I^+S^+ \right) \label{eq:effH_Pi1/2} ,
\end{eqnarray}
where $h_{1/2}=a - \frac{1}{2} (b+c)$, since $\Lambda=1$ and $\Sigma = -\frac{1}{2}$.

In the case of the $^2\Pi_{3/2}$ state, $\Lambda$-doubling can only arise from second order coupling with the $\Pi_{1/2}$ state. Again, due to the large splitting between the two states, this interaction will be extremely small, resulting in a vanishing $\Lambda$-doubling. Consequently, the term involving $d$ can be neglected and the Hamiltonian is simply:  
\begin{eqnarray}
H^{\text{F\&F}}_{\Pi_{3/2}}  &=& h_{3/2} I_z ,
\label{eq:effH_Pi3/2}
\end{eqnarray}
where $h_{3/2}=a + \frac{1}{2} (b+c)$, since $\Lambda=1$ and $\Sigma = \frac{1}{2}$.

Finally, expressions for the energy splitting between two hyperfine levels ($F=J \pm I$) can be determined by taking the expectation value of Eq. (\ref{eq:effH_Pi1/2}) and Eq.~(\ref{eq:effH_Pi3/2}) over the appropriate Hund's case a) functions~\cite{Dousmanis1955}:
\begin{eqnarray}
\Delta E_{^2\Pi_{1/2}}^{\text{F\&F}}  = \left[ h_{1/2} \pm \frac{1}{2}d(2J+1) \right]  \frac{2J + 1}{4J(J+1)}
\label{eq:deltaE12}
\end{eqnarray}
and 
\begin{eqnarray}
E_{^2\Pi_{3/2}}^{\text{F\&F}} = 3 h_{3/2} \frac{2J + 1}{4J(J+1)} .
\label{eq:deltaE32}
\end{eqnarray}

\subsection{Energy splittings in terms of the \textit{ab initio} parameters}\label{sec:ab_initio}

The $\mApar$ and $\mAperp$ constants introduced in Sec. \ref{sec:BaF_HFS_theory} (Eqs. (\ref{hyp_ff}), (\ref{eq:Apar}) and (\ref{eq:Aperp})), can be defined by the following effective Hamiltonian:
\begin{eqnarray}\label{eq:AparAperp}
H^{\textit{Ab initio}} &=& \mApar I_z \tilde{S}_z + \mAperp (I_x\tilde{S}_x + I_y\tilde{S}_y),
\end{eqnarray}
where $\vec{\tilde{S}}$ is the effective electronic spin. The Hamiltonian in Eq. (\ref{eq:AparAperp}) applies to all of the three considered states. 

In order to derive the energy splittings between the hyperfine levels, we again need to evaluate the expectation value of $H^{Ab~initio}$ over the appropriate spin functions.
The resulting energy splitting between two hyperfine levels ($F=J\pm I$) for a system with $I=1/2$ and $\Omega=1/2$ can be found in Ref. \cite{Sauer1999}:
\begin{eqnarray}
\Delta E_{\Omega=1/2}^{\textit{Ab initio}}(J) = \left(\frac{ \mApar}{2J+1} + \eta \mAperp \right)\frac{(2J+1)^2}{8J(J+1)}
\label{eq:Sauer}
\end{eqnarray}
where $\eta=-1$ for $e$ (even, symmetric) states and $\eta=+1$ for $f$ (odd, asymmetric) states. The corresponding parity is given by $p_e = (-1)^{J-1/2}$ and $p_f = - p_e$~\cite{Brown2003}. %p. 251.

For a $\Omega=3/2$ state, the parity-depending term vanishes due to the absence of $\Lambda$-doubling and the equation for the energy splitting takes the form:
\begin{eqnarray}
\Delta E_{\Omega=3/2}^{\textit{Ab initio}}(J) = \frac{9\mApar(2J+1)}{8J(J+1)}.
\label{eq:epsilon_3/2}
\end{eqnarray}

\subsection{Relations between \textit{ab initio} and Frosch \& Foley parameters}\label{sec:relations}

\begin{table}[t]
    \caption{Relations between the \textit{ab initio} and the F\&F parameters. }
    \label{tab:relations}
    \centering
    \begin{tabular}{lll}
    \hline
    Electronic state & \textit{ab initio} & Frosch \& Foley \\
    \hline
    $X^2\Sigma$ & $A_{\parallel}$       & $b + c$       \\
                & $A_{\perp}$           & $b$            \\
    $A^2\Pi$    & $A_{\parallel,1/2}$   & $2 h_{1/2}$     \\
                & $A_{\perp,1/2}$       & $-d$       \\
                & $A_{\parallel,3/2}$   & $\frac{2}{3} h_{3/2}$ \\
                \noalign{\smallskip}
    \hline
    \end{tabular}
\end{table}

Comparing the expressions for the energy splitting for the three states, we can derive the relations between the \textit{ab initio} and the Frosch \& Foley parameters;  these are shown in Table~\ref{tab:relations}. While there is one set ($\Apar$ and $\Aperp$) of \textit{ab initio} parameters for each electronic state, a single set of Frosch \& Foley parameters ($a$, $b$, $c$ and $d$) describes the entire $\Pi$ state manifold. 

\subsection{Analysis of experimental energy splittings}\label{sec:fit}

\begin{table}[b]
    \centering
    \caption{Experimental results of the energy \textit{splitting} between the hyperfine states $F=J+I$ and $F=J-I$ of the $A^2\Pi$ state taken from Ref. \cite{Ginny} together with the corresponding energy splittings and residuals obtained from fitting Eq.~(\ref{eq:deltaE12}) and (\ref{eq:deltaE32}) to this data.
    All results are in MHz. The fit parameters are given in Table~\ref{tab:FenF}.} 
    \label{tab:energies}
    \begin{tabular}{ccclrrl}
    \hline
     & $J$ & parity & $\Delta E$ measured & $\Delta E$ fitted & residual \\
    \hline
    $A^2\Pi_{1/2}$  & 1/2 &  f / -     & 21.89 $\pm$ 0.10 & 21.924 & -0.018 \\
                    &     &  e / +     & 16.65 $\pm$ 0.3  & 16.57 & 0.08 \\ \noalign{\smallskip}
                    & 3/2 &  f / +     & 10.23 $\pm$ 0.3  & 9.84 & 0.39 \\
    $A^2\Pi_{3/2}$  & 3/2 & e / -      & 19.27 $\pm$ 0.18 & 19.02 & 0.25 \\
                    &     & f / +      & 18.95 $\pm$ 0.13 & 19.02 & -0.07 \\
                    & 5/2 & f / -      & 12.17 $\pm$ 0.11 & 12.23 & -0.06 \\
    \hline
    \end{tabular}
\end{table}

\begin{table}[t]
    \caption{Frosch \& Foley parameters as deduced from experimental hyperfine splittings. Values in MHz.}
    \label{tab:FenF}
    \centering
    \begin{tabular}{lll}
    \hline
    Electronic state & ~~~~~~~~~~~~~~ & \\
    \hline
    $A^2\Pi$    & $a$                   & 26.33 $\pm$ 0.14    \\
                & $b+c$                 & -5.09 $\pm$ 0.29       \\
                & $d$                   & -4.01 $\pm$ 0.26 \\
                \noalign{\smallskip}
    \hline
    \end{tabular}
\end{table}

In order to find the experimental best values for $a$, $(b+c)$ and $d$ for the $\Pi$ state manifold, a weighted nonlinear least square fit of Eq.~(\ref{eq:deltaE12}) and Eq.~(\ref{eq:deltaE32}) was performed to the measured energy splittings shown in Table~\ref{tab:energies}. The results of this fit are given in Table~\ref{tab:FenF}. 
The small residuals presented in Table~\ref{tab:energies} show good agreement between derived equations and the measurements. 

\subsection{Comparison between theory and experiment} \label{sec:comparison}

In Table~\ref{tabfinal}, we present the experimentally determined values for the $A_{\parallel, 1/2}$, $A_{\perp, 1/2}$ and $A_{\parallel,3/2}$  
of the $\Pi$ manifold, derived from the fitted Frosch \& Foley parameters and the relations shown in Table~\ref{tab:relations}. These results are compared to the \textit{ab initio} values. For the $^2\Sigma$ ground state, the \textit{ab initio} values are compared to the experimental results of Ernst et al.~\cite{ernst1986}.

For the ground state, the CCSD(T) method was used, and the theoretical uncertainty was estimated to be around 3 \% and 2.5\% for $\Apar$ and $\Aperp$, respectively. The deviation from the experimental results are -0.7 and 0.8\% and thus well within the theoretical uncertainty estimate. This is similar to the accuracy we obtained in our earlier work for the HFS parameters of the same state, but for the $^{137}$Ba nucleus~\cite{Haase2020}. 

For the excited states, the FSCC method was used, resulting in much larger uncertainties, namely $\sim$17 and $\sim$31\% for $\Apar$ in $^2\Pi_{1/2}$ and $^2\Pi_{3/2}$, respectively, and $\sim$85\% for $\Aperp$ in $^2\Pi_{1/2}$. These large uncertainties are mostly due to the fact that the excited state parameters are much more sensitive to the basis set quality. In the case of $\Apar$ of both states, increasing the size of the basis set resulted in an increase of the values of $\Apar$, which means that the final results can be expected to underestimate the experimental values. This is indeed what we observe, i.e. the $\Apar$ is underestimated by 13.2 and 20.5\% for the $^2\Pi_{1/2}$ and $^2\Pi_{3/2}$ states, respectively. These deviations also fall within the estimated uncertainties. 

The basic idea of the theoretical uncertainty estimate is to account for all effects which are neglected in the \textit{ab initio} calculation, i.e. an agreement is to be expected between the theoretical and experimental values.
This is not the case for $\Aperp$ in the $^2\Pi_{1/2}$ state. 

As can be seen in Fig.~\ref{figerror}, the basis set effects dominate the uncertainty, being as large as $\sim$ 80\%. This indicates that the 4z quality basis set is insufficient for a proper description of the $\Aperp$ in the $^2\Pi_{1/2}$ state (and serves as further motivation for the development of Dyall's 5z quality basis sets \cite{Dyall2021}). Consequently, the uncertainty estimate of this value can be expected to be less reliable than for the other values presented here.

\begin{table}[t]  
\caption{Final recommended \textit{ab initio} values of $\Apar$ and $\Aperp$ of $^{19}$F in BaF with estimated uncertainties together with the values obtained from experiment. All theoretical results were obtained with the s-aug-cv4z basis set; the CCSD(T) method was used for the $^2\Sigma$ state and the FSCC method for the $\Pi$ states. All results are in MHz.}
    \label{tabfinal}
    \centering
    \begin{tabular}{llcrr}
    \hline
    State &  & \multicolumn{2}{c}{Theory } & \multicolumn{1}{c}{Experiment } \\
    \hline
    $^2\Sigma_{1/2}$ & $\Apar$  &  CCSD(T)  & $ 71.22\pm 2.14$  & ~~~71.73 $\pm$ 0.09$^a$ \\
                     & $\Aperp$ &"          & $64.02\pm 1.62$   & 63.51 $\pm$ 0.03$^a$ \\
    $^2\Pi_{1/2}$    & $\Apar$  & FSCC      & $ 50.13\pm 8.25$  & 57.7   $\pm$ 0.4$^b$\phantom{0}       \\
                     & $\Aperp$ &"          & $ 1.78 \pm 1.51 $ & 4.01   $\pm$ 0.26$^b$      \\
    $^2\Pi_{3/2}$    & $\Apar$  & FSCC      & $ 12.60\pm 3.86$  & 15.85  $\pm$ 0.14$^b$      \\
    \hline
    \multicolumn{5}{l}{\footnotesize{$^a$Reference \cite{ernst1986}}}\\
    \multicolumn{5}{l}{\footnotesize{$^b$Reference \cite{Ginny}}}
    \end{tabular}

\end{table}

\section{Conclusions}

We have carried out \textit{ab initio} calculations of the $^{19}$F HFS constants in the $^2\Sigma$ ground and the $^2\Pi_{1/2}$ and $^2\Pi_{3/2}$ excited states of BaF. For the ground state, the relativistic CCSD(T) method was used, while for the excited states, which require a multi-reference treatment, the relativistic FSCC (1,0) method was employed. 

A computational study of the effects of the basis set quality and the treatment of electron correlation on the calculated HFS constants was performed. We observed that the excited state HFS constants were much more sensitive to the basis set quality, compared to the ground state parameters. Furthermore, inclusion of perturbative triple excitations was found to have a very large effect on the calculated ground state HFS constants, much larger than that observed for the $^{137}$Ba coupling in our previous study \cite{Haase2020}.

The performed computational study was used to determine the uncertainties of the calculated values. These uncertainties are much smaller for the ground state parameters ($\sim$3\%) compared to the excited states ($\sim$16-85\%), mainly due to the larger basis set dependence of the latter. We also present the experimental HFS constants, obtained by fitting the measured energy splittings \cite{Ginny} and the relations between the \textit{ab initio} parameters and the Frosch \& Foley parameters, required to connect between experiment and theory. The \textit{ab initio} values agreed with the experimental results within the estimated uncertainties, except for the $\Aperp$ parameter of the $^2\Pi_{1/2}$ state, where higher quality basis sets are needed to produce accurate results. 

This work demonstrates the predictive power of the FSCC method and the reliability of the established uncertainty estimates. Uncertainty estimates based on purely theoretical considerations can be crucial in cases where the calculated property cannot be compared to experiment, and the findings of this work provide an important validation of our computational procedure.  

\section*{Acknowledgement}
We would like to thank the Center for Information Technology of the University of Groningen for their support and for providing access to the Peregrine high performance computing cluster.
% as well as M. G. Kozlov for helpful discussions
The NL-eEDM consortium receives program funding (EEDM-166) from the Netherlands Organisation for Scientific Research (NWO). The data that support the findings of this study are available from the corresponding author upon reasonable request.

\newpage
\bibliography{Pi.bib,mhypBaF.bib}
\end{document}